\def\be{\begin{equation}}
\def\ee{\end{equation}}
\def\bea{\begin{eqnarray}}
\def\eea{\end{eqnarray}}
\def\bq{\begin{quote}}
\def\eq{\end{quote}}
\def\bseq{\begin{subequation}}
\def\eseq{\end{subequation}}
\def\bsea{\begin{subeqnarray}}
\def\esea{\end{subeqnarray}}
\def\sb{\sin\beta}
\def\cb{\cos\beta}
\def\cbb{\cos 2 \beta}
\def\c2bb{\cos^2 2 \beta}

\def\s2bb{\sin^2 2 \beta}

\def\simgt{\stackrel{>}{{}_\sim}}
\def\ov{\overline}

\def\tb{\tan\beta}

\def\matb{(m_A,\tan \beta)}

\def\msta{m_{\tilde{t}_1}}
\def\mstb{m_{\tilde{t}_2}}

\def\st{\tilde{t}}

\def\tev{\rm \; TeV}
\def\gev{{\rm \; GeV}}

\def\tsapp{\theta^*_{\rm app}}
\documentstyle[12pt]{article}
\catcode`@=11
\def\marginnote#1{}
\newcount\hour
\newcount\minute
\newtoks\amorpm
\hour=\time\divide\hour by60
\minute=\time{\multiply\hour by60 \global\advance\minute by-\hour}
\edef\standardtime{{\ifnum\hour<12 \global\amorpm={am}%
        \else\global\amorpm={pm}\advance\hour by-12 \fi
        \ifnum\hour=0 \hour=12 \fi
        \number\hour:\ifnum\minute<10 0\fi\number\minute\the\amorpm}}
\edef\militarytime{\number\hour:\ifnum\minute<10 0\fi\number\minute}
\def\draftlabel#1{{\@bsphack\if@filesw {\let\thepage\relax
   \xdef\@gtempa{\write\@auxout{\string
      \newlabel{#1}{{\@currentlabel}{\thepage}}}}}\@gtempa
   \if@nobreak \ifvmode\nobreak\fi\fi\fi\@esphack}
        \gdef\@eqnlabel{#1}}
\def\@eqnlabel{}
\def\@vacuum{}
\def\draftmarginnote#1{\marginpar{\raggedright\scriptsize\tt#1}}
\def\draft{\oddsidemargin 0.0truein
        \def\@oddfoot{\sl preliminary draft \hfil
        \rm\thepage\hfil\sl\today\quad\militarytime}
        \let\@evenfoot\@oddfoot \overfullrule 3pt
        \let\label=\draftlabel
        \let\marginnote=\draftmarginnote
   \def\@eqnnum{(\theequation)\rlap{\kern\marginparsep\tt\@eqnlabel}%
\global\let\@eqnlabel\@vacuum}  }
\catcode`@=12
\begin{document}
\begin{titlepage}
\vspace*{-1cm}
\noindent
\hfill{hep-ph/9312296}
\\
\phantom{bla}
\hfill{CERN-TH.7057/93}
\\
\phantom{bla}
\hfill{IEM-FT-81/93}
\\
\phantom{bla}
\hfill{FTUAM-36/93}
\vskip 1.5cm
\begin{center}
{\Large\bf Aspects  of the electroweak phase transition}
\end{center}
\begin{center}
{\Large\bf in the Minimal Supersymmetric Standard Model}
\footnote{Work supported in part by the European Union under contract
No.~CHRX-CT92-0004.}
\end{center}
\vskip 0.5cm
\begin{center}
{\large A. Brignole} \footnote{Supported by a postdoctoral fellowship
of the Ministerio de Educaci\'on y Ciencia, Spain. Address after
November
7, 1993: Theory Group, Lawrence Berkeley Laboratory, Berkeley CA
94720, USA.},
\\
Dep. de F\'{\i}sica Te\'orica, Universidad Aut\'onoma de Madrid\\
Cantoblanco, E-28049 Madrid, Spain
\vskip .3cm
{\large J.R. Espinosa} \footnote{Supported by a grant of Comunidad
de Madrid, Spain.},
{\large M. Quir\'os} \footnote{Work partly supported by CICYT, Spain,
under contract AEN93-0139.}
\\
Instituto de Estructura de la Materia, CSIC\\
Serrano 123, E-28006 Madrid, Spain\\
\vskip .2cm
and \\
\vskip .2cm
{\large F. Zwirner} \footnote{On leave from INFN, Sezione di Padova,
Padua, Italy.}
\\
Theory Division, CERN \\
CH-1211 Geneva 23, Switzerland \\
\end{center}
\vskip .4cm
\begin{abstract}
\noindent
We study the finite-temperature effective potential of the Minimal
Supersymmetric Standard Model in the full $(m_A,\tan\beta)$ parameter
space. As for the features of the electroweak phase transition, we
identify two possible sources of significant differences with respect
to the Standard Model: a stop sector with little supersymmetry
breaking
makes the phase transition more strongly first-order, whereas a
light
CP-odd neutral boson weakens its first-order nature. After
including the
leading plasma effects, $T=0$ radiative corrections due to top and
stop
loops, and the most important experimental constraints, we find that
the
danger  of washing out any baryon asymmetry created at the
electroweak scale
is in general no less than in the Standard Model.
\end{abstract}
\vfill{
CERN-TH.7057/93
\newline
\noindent
December 1993}
\end{titlepage}
\setcounter{footnote}{0}
\newpage

{\bf 1.}
The option of generating the cosmological baryon asymmetry at the
electroweak phase transition is not necessarily the one chosen by
Nature, but it is certainly fascinating, and has recently deserved a
lot
of attention [\ref{reviews}]. At the qualitative level, the Standard
Model (SM) meets the basic requirements for a successful
implementation
of this scenario. At the quantitative level, however, it suffers from
two basic problems. First, the amount of CP violation in the
Kobayashi-Maskawa matrix appears to be
insufficient, even after taking into account the large uncertainties
associated with dynamical models of baryogenesis\footnote{Unorthodox
views on this point, recently put forward in [\ref{farsha}], have
been
subsequently  questioned in [\ref{gavela}].}. Secondly, one must make
sure that sphaleron interactions in the broken phase do not
wash out,
at the completion of the phase transition, any previously created
baryon
asymmetry. This requirement suggests that the transition should be
rather
strongly first order, $v(T_C)/T_C \simgt 1$, where $T_C$ is the
critical
temperature and $v(T_C)$ is the symmetry-breaking vacuum expectation
value.
In the SM, this condition turns out to be incompatible with the
experimental
limits [\ref{coignet}] on the Higgs mass, $m_{\phi} > 63.5$ GeV  at
$95 \%$
c.l., even after implementing the conventional techniques for dealing
with
the infrared problem [\ref{reviews},\ref{limits}]. To cure both
problems,
one can consider plausible extensions of the SM, to see if they can
allow
for additional sources of CP violation and for an enhanced strength
of the
first-order phase transition. Among these extensions, the physically
most motivated and phenomenologically most acceptable one is the
Minimal
Supersymmetric Standard Model (MSSM). This model allows for extra
CP-violating phases besides the Kobayashi-Maskawa one, which could
help
in generating the observed baryon asymmetry [\ref{cn}]. It is then
interesting to study whether in the MSSM the nature of the phase
transition can be significantly modified with respect to the SM.

In a recent paper [\ref{eqz3}], some of us have considered the MSSM
in the
limit $m_A \rightarrow \infty$, corresponding to
only
one light Higgs with SM properties, and improved over previous
studies
[\ref{previous}] by including a full discussion of the top/stop
sector,
and by resumming the leading plasma corrections to gauge boson and
stop
masses. It was found that this special limit of the MSSM can only
marginally
improve the situation with respect to the SM case. In the present
paper, we
extend the considerations of [\ref{eqz3}] to the full $\matb$
parameter space,
characterizing the Higgs sector of the MSSM. Even barring the
interesting
possibility of spontaneous CP-violation at finite temperature
[\ref{cpspo}],
as well as the possibility of charge- and colour-breaking minima, we
have to
deal with a complicated two-variable potential, which requires a
numerical
analysis. However, to allow for an understanding of the
behaviour of various quantities,
we also produce some approximate analytical formulae. After
including the most important experimental constraints, we find that
there is very little room for the MSSM to improve over the SM.

\vspace{1cm}
{\bf 2.}
The main tool for our study is the one-loop, daisy-improved
finite-temperature
effective potential of the MSSM, $V_{\rm{eff}}(\phi,T)$. We are
actually
interested in the dependence of the potential on $\phi_1 \equiv {\rm
Re}
\, H_1^0$ and $\phi_2 \equiv {\rm Re} \,  H_2^0$ only, where $H_1^0$
and $H_2^0$ are the neutral
components of the Higgs doublets $H_1$ and $H_2$, thus $\phi$ will
stand
for $(\phi_1,\phi_2)$. Working in the 't~Hooft-Landau gauge and in
the
$\ov{DR}$-scheme, we can write
\be
\label{total}
V_{\rm{eff}}(\phi,T) = V_0(\phi)
+ V_1(\phi,0) + \Delta V_1(\phi,T)
+\Delta V_{\rm{daisy}}(\phi,T) \, ,
\ee
where
\bea
\label{v0}
V_0(\phi) & = &  m_1^2 \phi_1^2 + m_2^2  \phi_2^2 + 2 m_3^2 \phi_1
\phi_2
+ {g^2+g'\,^2 \over 8} (\phi_1^2 -\phi_2^2)^2 \, ,
\\
\label{deltav}
V_1(\phi,0) & = & \sum_i {n_i \over
64 \pi^2} m_i^4 (\phi) \left[ \log {m_i^2 (\phi) \over Q^2} - {3
\over 2} \right]  \, ,
\\
\label{deltavt}
\Delta V_1(\phi,T) & = & {T^4 \over 2 \pi^2} \left\{ \sum_i
n_i \, J_i \left[ { m^2_i (\phi) \over T^2 } \right]  \right\} \, ,
\\
\label{dvdaisy}
\Delta V_{\rm{daisy}}(\phi,T) & = & - {T \over 12 \pi} \sum_i n_i
\left[ \ov{m}_i^3 (\phi, T ) - m_i^3 (\phi) \right] \, .
\eea
The four contributions (\ref{v0}--\ref{dvdaisy}) to the effective
potential
(\ref{total}) have the following meaning. The first term,
eq.~(\ref{v0}),
is the tree-level potential. The second term, eq.~(\ref{deltav}), is
the one-loop contribution at $T=0$: $Q$ is the renormalization scale,
where we
choose for
definiteness $Q^2 = m_Z^2$, $m_i^2 (\phi)$ is the field-dependent
mass of
the $i^{th}$ particle, and $n_i$ is the corresponding number of
degrees of
freedom, taken negative for fermions. Since $V_1(\phi,0)$ is
dominated by
top ($t$) and stop ($\st_1,\st_2$) contributions, only these will be
included
in the following. The third term, eq.~(\ref{deltavt}), is the
additional
one-loop
contribution due to temperature effects. Here $J_i=J_+ (J_-)$ if the
$i^{th}$
particle is a boson (fermion), and
\be
\label{ypsilon}
J_{\pm} (y^2) \equiv \int_0^{\infty} dx \, x^2 \,
\log \left( 1 \mp e^{- \sqrt{x^2 + y^2}} \right) \, .
\ee
Since the relevant contributions to $\Delta V_1(\phi,T)$ are due to
top ($t$),
stops ($\st_1,\st_2$) and gauge bosons ($W,Z$), only these will be
considered
in the following. Finally, the last term, eq.~(\ref{dvdaisy}), is a
correction
coming from the resummation of the leading infrared-dominated
higher-loop
contributions, associated with the so-called daisy diagrams. The sum
runs over bosons only. The masses $\ov{m}_i^2 (\phi,T)$ are obtained
from the ${m}_i^2 (\phi)$ by adding the leading $T$-dependent
self-energy
contributions, which are proportional to $T^2$. We recall that, in
the gauge
boson sector, only the longitudinal components ($W_L, Z_L, \gamma_L$)
receive
such contributions.

The relevant degrees of freedom for our calculation are:
\be
\label{multi}
n_t = - 12 \, ,  \;\;
n_{\st_1} = n_{\st_2} = 6 \,  , \;\;
n_W=6 \,  , \;\; n_Z=3 \,  , \;\;
n_{W_L}=2 \, ,  \;\; n_{Z_L}=n_{\gamma_L}=1 \, .
\ee
The field-dependent top mass is
\be
\label{tmass}
m_t^2(\phi)=h_t^2 \phi_2^2 \, .
\ee
The entries of the field-dependent stop mass matrix are
\bea
\label{tlmass}
m_{\tilde{t}_L}^2 (\phi) & = & m_{Q_3}^2 + m_t^2 (\phi) +
D_{\tilde{t}_L}^2 (\phi)  \, ,
\\
\label{trmass}
m_{\tilde{t}_R}^2 (\phi) & = & m_{U_3}^2 + m_t^2 (\phi) +
D_{\tilde{t}_R}^2 (\phi)  \, ,
\\
\label{mixmass}
m_X^2 (\phi) & = & h_t (A_t \phi_2 + \mu \phi_1) \, ,
\eea
where $m_{Q_3}$, $m_{U_3}$ and $A_t$ are soft supersymmetry-breaking
mass
parameters, $\mu$ is a superpotential Higgs mass term, and
\bea
\label{dterms}
D_{\tilde{t}_L}^2(\phi)& = & \left( {1 \over 2}-
{2 \over 3}\sin^2 \theta_W \right)
{g^2 + g'\,^2 \over 2}(\phi_1^2-\phi_2^2),\\
D_{\tilde{t}_R}^2(\phi) &=&\left( {2 \over 3}\sin^2 \theta_W \right)
{g^2 + g'\,^2 \over 2}(\phi_1^2-\phi_2^2)
\eea
are the $D$-term contributions. The field-dependent stop masses are
then
\be
\label{mstop}
m_{\tilde{t}_{1,2}}^2 (\phi) = {m^2_{\tilde{t}_L} (\phi) +
m^2_{\tilde{t}_R} (\phi) \over 2} \pm \sqrt{ \left[
{m^2_{\tilde{t}_L} (\phi)- m^2_{\tilde{t}_R} (\phi) \over 2}
\right]^2 + \left[ m_X^2(\phi) \right]^2  } \, .
\ee
The corresponding effective $T$-dependent masses,
$\ov{m}^2_{\tilde{t}_{1,2}}
(\phi,T)$, are given by expressions identical to (\ref{mstop}), apart
from the
replacement
\be
\label{repl}
m^2_{\tilde{t}_{L,R}} (\phi) \, \rightarrow \,
\ov{m}^2_{\tilde{t}_{L,R}}(\phi,T) \equiv
m^2_{\tilde{t}_{L,R}} (\phi)+  \Pi_{\tilde{t}_{L,R}}(T)  \,  .
\ee
The $\Pi_{\tilde{t}_{L,R}}(T)$ are the leading parts of the
$T$-dependent
self-energies of $\tilde{t}_{L,R}$\,,
\bea
\label{pistl}
\Pi_{\tilde{t}_L}(T)& = &
{4 \over 9}g_s^2 T^2 +
{1 \over 4}g^2 T^2 +
{1 \over 108}g'\,^2  T^2 +
{1 \over 6}h_t^2 T^2 \, ,
\\
\label{pistr}
\Pi_{\tilde{t}_R}(T) & =  &
{4 \over 9}g_s^2 T^2 +
{4 \over 27} g'\,^2 T^2 +
{1 \over 3}h_t^2 T^2 \, ,
\eea
where $g_s$ is the strong gauge coupling constant. Only loops of
gauge
bosons, Higgs bosons and third generation squarks have been included,
implicitly assuming that all remaining supersymmetric particles are
heavy and decouple. If some of these are also light, the plasma
masses for the stops  will be even larger, further suppressing the
effects of the associated cubic terms, and therefore weakening the
first-order nature of the phase transition. Finally, the
field-dependent
gauge boson masses are
\be
\label{gauge}
m_W^2 (\phi) = {g^2 \over 2} (\phi_1^2 + \phi_2^2)  \, ,
\;\;\;\;\;\;
m_Z^2 (\phi) = {g^2 + g'\,^2 \over 2} (\phi_1^2 + \phi_2^2)  \, ,
\ee
and the effective $T$-dependent masses of the longitudinal gauge
bosons
are
\bea
\label{wmass}
\phantom{b}&\phantom{b}&
\ov{m}_{W_L}^2(\phi,T) \ \  =  m_W^2 (\phi)+ \Pi_{W_L}(T)  \, ,
\\
\label{zphlmass}
\phantom{b}&\phantom{b}&
\ov{m}^2_{Z_L,\gamma_L}(\phi,T)  =
\frac{1}{2} \left[ m_Z^2 (\phi) + \Pi_{W_L}(T) + \Pi_{B_L}(T) \right]
\nonumber\\
& \pm &
\sqrt{ \frac{1}{4} \left[ {g^2 - g'\,^2 \over 2} (\phi_1^2 +
\phi_2^2)
+ \Pi_{W_L}(T) - \Pi_{B_L}(T) \right]^2
+ \left[ {gg' \over 2}(\phi_1^2 + \phi_2^2) \right]^2 } \, .
\eea
In eqs.~(\ref{wmass}) and (\ref{zphlmass}), $\Pi_{W_L}(T)$ and
$\Pi_{B_L}(T)$
are the leading parts of the $T$-dependent self-energies of $W_L$ and
$B_L$,
given by
\be
\label{piwb}
\Pi_{W_L}(T)  =  {5 \over 2}g^2 T^2 \, ,
\;\;\;\;\;
\Pi_{B_L}(T) =   {47 \over 18}g'\,^2 T^2 \, ,
\ee
where only loops of Higgs bosons, gauge bosons, Standard Model
fermions
and third-generation squarks have been included.

\vspace{1cm}
{\bf 3.}
We shall now analyse the effective potential (\ref{total}) as a
function of
$\phi$ and $T$. Before doing this, however, we trade the parameters
$m_1^2,
m_2^2,m_3^2$ appearing in the tree-level potential (\ref{v0}) for
more
convenient parameters. To this purpose, we first minimize the
zero-temperature
effective potential, i.e. we impose the vanishing of the first
derivatives of
$V_0(\phi) + V_1(\phi,0)$ at $(\phi_1,\phi_2)=(v_1,v_2)$, where
$(v_1,v_2)$
are the one-loop vacuum expectation values at $T=0$. This allows us
to eliminate
$m_1^2$ and $m_2^2$ in favour of $m_Z^2$ and $\tan\beta\equiv
v_2/v_1$:
\bea
\label{12away}
m_1^2 & = & -m_3^2 \tan\beta -{m_Z^2 \over 2} \cos 2\beta - \sum_i
{n_i
\over 64 \pi^2} \left[ {\partial  m_i^2  \over  \partial \phi_1}
{m_i^2 \over \phi_1} \left( \log {m_i^2 \over Q^2} - 1 \right)
\right]_{\phi_{1,2}=v_{1,2}}  \, ,
\\
m_2^2 & = & -m_3^2 \cot\beta +{m_Z^2 \over 2} \cos 2\beta - \sum_i
{n_i
\over 64 \pi^2} \left[ {\partial  m_i^2 \over \partial \phi_2}
{m_i^2 \over \phi_2} \left( \log {m_i^2 \over Q^2} - 1 \right)
\right]_{\phi_{1,2}=v_{1,2}} \, .
\eea
Moreover, $m_3^2$ can be traded for the one-loop-corrected mass
$m_A^2$
of the CP-odd neutral Higgs boson. In our approximation [\ref{erz2}]
\be
\label{3away}
m_3^2 = - m_A^2 \sb\cb- \frac{3g^2 m_t^2 \mu A_t}{32\pi^2 m_W^2
\sin^2\beta}
\frac{\msta^2 \left( \log {\msta^2 \over Q^2} - 1 \right) -\mstb^2
\left(
\log {\mstb^2 \over Q^2} - 1 \right)} {\msta^2-\mstb^2} \, .
\ee
Therefore the whole effective potential (\ref{total}) is completely
determined, in our approximation, by the parameters $(m_A,\tb)$ of
the Higgs sector, and by the parameters ($m_t$, $m_{Q_3}$, $m_{U_3}$,
$\mu$, $A_t$) of the top/stop sector. The same set of parameters also
determines the one-loop-corrected masses and couplings of the MSSM
Higgs bosons.

The next steps are the computation of the critical temperature and
of the location of the minimum of the effective potential at the
critical temperature.
We define here $T_0$ as the temperature at which the determinant of
the
second derivatives of $V_{\rm eff}(\phi,T)$ at $\phi=0$ vanishes:
\be
\label{det}
\det\left[
{{\partial^2 V_{\rm eff}(\phi,T_0)}\over{\partial \phi_i \partial
\phi_j}}
\right]_{\phi_{1,2}=0} = 0  \, .
\ee
It is straightforward to compute the derivatives in eq.~(\ref{det})
from the previous formulae; the explicit expressions are
\bea
\label{secder}
\frac{1}{2} \left[
{{\partial^2 V_{\rm eff}}\over{\partial \phi_i^2 }}\right]_0
& = &
m_i^2
+{1 \over 64 \pi^2}
\left[6a_{ii}  m_{Q_3}^2  \left( \log {m_{Q_3}^2 \over Q^2} - 1
\right)
+6b_{ii} m_{U_3}^2  \left( \log {m_{U_3}^2 \over Q^2} - 1
\right)\right]
\nonumber\\
& & +{T^2 \over 4 \pi^2}
\left[ {\pi^2\over 12}(9g^2+3g'\,^2+\delta_{i2} \cdot 12h_t^2)
+6a_{ii} J'_+ \left({ m_{Q_3}^2  \over T^2}  \right)
+6b_{ii} J'_+ \left({ m_{U_3}^2  \over T^2} \right)\right]
\nonumber\\
& & -{T \over 16 \pi}
\left\{ 3g^2 \left[ \Pi_{W_L}(T) \right]^{\frac{1}{2}}
+g'\,^2 \left[ \Pi_{B_L}(T) \right]^{\frac{1}{2}} \right.
\nonumber\\
& &   +6 \left[ \ov{a}_{ii}\left(m_{Q_3}^2+
\Pi_{\tilde{t}_L}(T) \right)^{\frac{1}{2}}-
{a}_{ii}\left(m_{Q_3}^2\right)^{\frac{1}{2}} \right]
\nonumber\\
& &  \left.
 +6 \left[ \ov{b}_{ii}\left( m_{U_3}^2+
\Pi_{\tilde{t}_R}(T) \right)^{\frac{1}{2}}-
{b}_{ii}\left(m_{U_3}^2\right)^{\frac{1}{2}} \right]  \right\} \, ,
\nonumber\\
\frac{1}{2} \left[
{{\partial^2 V_{\rm eff}}\over{\partial \phi_1 \partial
\phi_2}}\right]_0
& = &
m_3^2
+{1 \over 64 \pi^2}
6a_{12}\left[m_{Q_3}^2  \left( \log {m_{Q_3}^2 \over Q^2} - 1 \right)
- m_{U_3}^2  \left( \log {m_{U_3}^2 \over Q^2} - 1 \right)\right]
\nonumber\\
& & +{T^2 \over 4 \pi^2} 6a_{12}
\left[J'_+ \left({ m_{Q_3}^2  \over T^2 } \right)
- J'_+ \left({ m_{U_3}^2  \over T^2 } \right)\right]
\nonumber\\
& & -{T \over 16 \pi}
\left\{ 6\ov{a}_{12}
\left[ \left(m_{Q_3}^2+\Pi_{\tilde{t}_L}(T) \right)^{\frac{1}{2}}-
\left(m_{U_3}^2+\Pi_{\tilde{t}_R}(T) \right)^{\frac{1}{2}}\right]
\right.
\nonumber\\
& &
\left. -6a_{12} \left[ \left( m_{Q_3}^2\right)^{\frac{1}{2}}
-\left(m_{U_3}^2\right)^{\frac{1}{2}} \right]  \right\} \, .
\eea
The coefficients $a_{ij},b_{ij}$ are given by
\bea
\label{abij}
a_{11} & \equiv &
\left( {1 \over 2}-{2 \over 3}\sin^2 \theta_W \right)(g^2 + g'\,^2)
+\frac{2h_t^2\mu^2}{m_{Q_3}^2-m_{U_3}^2} \, ,
\nonumber\\
b_{11} & \equiv &
\left({2 \over 3}\sin^2 \theta_W \right)(g^2 + g'\,^2)
-\frac{2h_t^2\mu^2}{m_{Q_3}^2-m_{U_3}^2} \, ,
\nonumber\\
a_{22} & \equiv &
2 h_t^2 - \left( {1 \over 2}-{2 \over 3}\sin^2 \theta_W \right)(g^2 +
g'\,^2)
+\frac{2h_t^2 A_t^2}{m_{Q_3}^2-m_{U_3}^2}
\nonumber\\b_{22} & \equiv &
2 h_t^2 - \left( {2 \over 3}\sin^2 \theta_W \right)(g^2 + g'\,^2)
-\frac{2h_t^2 A_t^2}{m_{Q_3}^2-m_{U_3}^2} \, ,
\nonumber\\
a_{12} & \equiv &
\frac{2h_t^2\mu A_t}{m_{Q_3}^2-m_{U_3}^2} \, ,
\eea
and the coefficients $\ov{a}_{ij},\ov{b}_{ij}$ are given
by identical expressions, apart from the replacement
\be
\label{replace}
m_{Q_3}^2-m_{U_3}^2 \, \rightarrow \,
m_{Q_3}^2-m_{U_3}^2 +
\Pi_{\tilde{t}_{L}}(T)-\Pi_{\tilde{t}_{R}}(T)    \,  .
\ee

Once eq.~(\ref{det}) is solved (numerically) and $T_0$ is found, one
can
minimize (numerically) the potential $V_{\rm eff}(\phi,T_0)$ and find
the
minimum $[v_1(T_0),v_2(T_0)]$. The quantity of interest is indeed, as
will
be discussed later, the ratio $v(T_0)/T_0$, where $v(T_0)\equiv
\sqrt{v_1^2
(T_0)+v_2^2(T_0)}$.

\vspace{1cm}
{\bf 4.}
Before moving to the discussion of our numerical results, we would
like to present some approximate analytical formulae, which will be
useful
for a qualitative understanding of the various dependences of $T_0$,
$v_1(T_0)$ and $v_2(T_0)$. In this paragraph, we shall work in the
limit of
heavy degenerate stops, $m_{Q_3} = m_{U_3} \equiv \tilde{m} \gg T_0$
and $A_t = \mu = 0$, neglecting the D-term contributions to the stop
squark masses, and keeping only the most important terms in the
high-temperature expansions of the $J_i$ functions for the gauge
bosons and the top quark. In the chosen limit, the effective
potential of eq.~(\ref{total}) can be approximately written, in the
polar  coordinates $\varphi \equiv \sqrt{\phi_1^2 + \phi_2^2}$ and
$\tan
\theta \equiv \phi_2 / \phi_1$, as
\be
\label{polar}
V_{\rm eff} (\phi,T) \simeq \left[ a(\theta) T^2 - b (\theta) \right]
\varphi^2 - E T \varphi^3 + {1 \over 4} \lambda_T ( \theta )
\varphi^4 \, ,
\ee
where
\bea
a (\theta) & = & {3 g^2 + g' \, ^2 \over 16} + {h_t^2 \over 4}
\sin^2 \theta \, , \\
b (\theta) & = & {m_Z^2 \over 2} \cos 2 \beta \cos 2 \theta -
m_A^2 \sin^2 ( \beta - \theta ) +
{3 h_t^2 \over 8 \pi^2} \sin^2 \theta \,  m_t^2
\left( 1 + \log {\tilde{m}^2 \over m_t^2} \right)
\, , \\
E & = & {2 \over 3} {\sqrt{2} \over 16 \pi}
\left[ 2 g^3 + (g^2 + g' \, ^2)^{3/2} \right] \, , \\
\label{elleti}
\lambda_T (\theta) & = & {1 \over 2}
(g^2 + g' \, ^2) \cos^2 2 \theta +
{3 h_t^4 \over 4 \pi^2} \sin^4 \theta \left(
\log {\tilde{m}^2 \over T^2} - 1.14 \right) \, .
\eea
We have exploited the fact that, in the high-temperature limit, the
terms proportional to $m_i^4 \log m_i^2$ cancel in the sum $V_1
(\phi,
0) + \Delta V_1 (\phi, T)$ (in the chosen limit, of course, we cannot
perform the high-temperature expansion on the stop contributions).

We define the critical angle $\theta^*$ by the flat direction of the
effective potential (\ref{total}) around $\varphi=0$ at $T=T_0$, and
we
denote by
$\tsapp$ the analogous quantity evaluated from the approximate
parametrization of eq.~(\ref{polar}). The critical temperature $T_0$
and the critical angle
$\tsapp$ are then determined by the conditions
\be
\label{condi}
\left\{
\begin{array}{l}
a(\tsapp) T_0^2 - b (\tsapp) = 0  \\
a'(\tsapp) T_0^2 - b' (\tsapp) = 0
\end{array}
\, .
\right.
\ee
Solving eq.~(\ref{condi}) amounts to solving eq.~(\ref{det}) and
finding
the eigenvector corresponding to the zero eigenvalue. One finds
\be
\label{T0}
T_0^2 = {- B + \sqrt{B^2 + A C} \over  A} \, ,
\ee
where
\bea
A & = & {3 g^2 + g' \, ^2 \over 16} \,
{3 g^2 + g' \, ^2 + 4 h_t^2 \over 16} \, , \\
B & = & {3 g^2 + g' \, ^2 + 2 h_t^2  (1 - \cos 2 \beta) \over 32}
m_A^2 - {h_t^2 \over 16} \cos 2 \beta m_Z^2
\nonumber  \\
& & -
{3 g^2 + g' \, ^2 \over 16} {3 h_t^2 \over 16 \pi^2}
m_t^2 \left( 1 + \log {\tilde{m}^2 \over m_t^2} \right)
\, , \\
C & = &
\left[ m_Z^2 \c2bb + (1 - \cbb) {3 h_t^2 \over 8 \pi^2}
m_t^2 \left( 1 + \log {\tilde{m}^2 \over m_t^2} \right) \right]
{m_A^2 \over 2}
\nonumber  \\
& & +
\left[ {m_Z^2 \over 2} \c2bb  - \cbb {3 h_t^2 \over 8 \pi^2}
m_t^2 \left( 1 + \log {\tilde{m}^2 \over m_t^2} \right) \right]
{m_Z^2 \over 2}
\, ,
\eea
and
\be
\label{tapp}
\tan 2 \theta^*_{\rm app} = {\displaystyle \tan 2\beta
\frac{m_A^2}{m_A^2+m_Z^2+\left[\frac{h_t^2 T_0^2}{4}
-\frac{3h_t^2}{8\pi^2}m_t^2\left(1+\log \frac{\tilde{m}^2}
{m_t^2}\right)\right]/\cos 2\beta } } \, ,
\ee
where $T_0$ is determined by eq.~(\ref{T0}).
If we now assume that $\tan \theta(T_0) \equiv v_2(T_0) /  v_1(T_0)$
can be approximated by $\tan \tsapp$, we can also write
\be
\label{voverT}
\left[ {v(T_0) \over T_0}\right]_{\rm app}
\simeq {3 E \over \lambda_{T_0} (\tsapp)} \, .
\ee

In fig.~1, we plot $\tan\theta(T_0)$ (solid lines) and $\tan\theta^*$
(dashed lines) as functions of $m_A$, for $\tan \beta = 1,2,5$ and
the representative parameter choice $m_t = 150 \gev$, $m_{Q_3} =
m_{U_3} = 1 \tev$, $A_t = \mu = 0$. We can see that, in all cases,
$\tan \theta (T_0) \simeq \tan \theta^*$ to a very good accuracy.
To check our analytical approximation, we also plot $\tan \tsapp$
(dotted lines) as obtained from eq.~(\ref{tapp}). We can see from
fig.~1 that all three quantities tend to the corresponding value of
$\tan \beta$ for large values of
$m_A$, whereas they increase to large values for small values of
$m_A$. This fact is a general trend of $V_{\rm eff}$, for arbitrary
values of the top/stop parameters. As for $v(T_0)/T_0$, we have also
checked that the analytical expression (\ref{voverT}) is an adequate
approximation to the numerical value obtained from (\ref{total}) and
(\ref{det}). The qualitative behaviour of $v(T_0)/T_0$ can then be
derived from (\ref{voverT}), (\ref{elleti}), (\ref{T0}) and
(\ref{tapp}): for fixed $m_A$, $v(T_0)/T_0$ increases when $\tb$ is
approaching 1 from above; for fixed $\tb$, $v(T_0)/T_0$ is an
increasing function of $m_A$.

\vspace{1cm}
{\bf 5.}
We now discuss the particle physics constraints on the parameters
of the top/stop sector and of the Higgs sector. To be as general as
possible,
we treat $m_{Q_3}$, $m_{U_3}$ and the other soft mass terms as
independent parameters, even if they can be related in specific
supergravity
models.

The constraints on the top/stop sector have already been discussed in
[\ref{eqz3}], so we just recall them briefly.  Direct and indirect
searches
at LEP [\ref{coignet}] imply that $m_{\tilde{b}_L} \simgt 45 \gev$,
which
in turn translates into a  bound in the $(m_{Q_3},\tb)$ plane.
Electroweak
precision measurements [\ref{altarelli}] put stringent constraints on
a light
stop-sbottom sector: in first approximation, and taking into account
possible
effects [\ref{abc}] of other light particles of the MSSM, we
conservatively
summarize the constraints by $\Delta \rho (t,b) + \Delta \rho
(\tilde{t},
\tilde{b}) < 0.01$, where the explicit expression for $\Delta \rho
(\tilde{t},
\tilde{b})$ can be found in [\ref{rhosusy}].

We finally need to consider the constraints coming from LEP searches
for
supersymmetric Higgs bosons [\ref{coignet}].
Experimentalists put limits on the processes $Z \rightarrow h Z^*$
and
$Z \rightarrow h A$, where $h$ is the lighter neutral CP-even boson.
We
need to translate these limits into exclusion contours in the $\matb$
plane,
for given values of the top/stop parameters. In order to do this, we
identify
the value of $BR(Z \rightarrow h Z^*)$, which corresponds to the
limit
$m_{\phi} > 63.5 \gev$ on the SM Higgs, and the value of $BR(Z
\rightarrow
h A)$, which best fits the published limits for the representative
parameter
choice $m_t = 140 \gev$, $m_{Q_3} = m_{U_3} \equiv \tilde{m} = 1
\tev$,
$A_t = \mu = 0$. We then compare those values of $BR(Z \rightarrow h
Z^*)$
and $BR(Z \rightarrow h A)$ with the theoretical predictions of the
MSSM,
for any desired parameter choice and after including the radiative
corrections
associated to top/stop loops  [\ref{pioneer},\ref{erz2}]. Of course,
this procedure is not entirely  correct, since it ignores the
variations of
the efficiencies  with the Higgs masses and branching ratios, as well
as
the possible presence of candidate events at some mass values, but it
is adequate for our purposes.

\vspace{1cm}
{\bf 6.}
We now present our numerical results, based on the effective
potential of eq.~(\ref{total}), concerning the strength of the
electroweak phase transition and the condition for preserving the
baryon asymmetry.
According to [\ref{shapo}], the condition to avoid  erasing any
previously
generated baryon asymmetry via sphaleron transitions is
\be
{E_{sph} (T_C) \over T_C} > 45 \, ,
\ee
where $T_C$ is the actual temperature at which the phase transition
occurs, satisfying the inequalities
\be
T_0 < T_C < T_D \, ,
\ee
if $T_0$ is defined by (\ref{det}) and $T_D$ is the temperature at
which
there are two degenerate minima. Particularizing to the MSSM the
studies
of sphalerons in general two-Higgs models [\ref{two}], we obtain
that
\be
E_{sph}^{MSSM} (T) \le E_{sph}^{SM} (T) \, ,
\ee
where, in our conventions,
\be
\label{esph}
{E_{sph}^{SM} (T) \over T} = {4 \sqrt{2} \pi \over g} B
\left\{ {\lambda_{\rm eff} [\theta(T)] \over 4 g^2 } \right\} {v(T)
\over T} \, ,
\ee
and $B$ is a smoothly varying function whose values can be found in
[\ref{km}]. For example, $B(10^{-2})=1.67$, $B(10^{-1})=1.83$, $B(1)
=2.10$. It can also be shown that
\be
{v(T_D) \over T_D} < {v(T_C) \over T_C} <
{v(T_0) \over T_0} \, .
\ee
Finally, the corrections in $E_{sph}^{SM}$ due to $g'\neq 0$
have been estimated and shown to be
small [\ref{mixing}]. Therefore, a conservative bound to be imposed
is
\be
\label{erre}
R \equiv {v(T_0) \over T_0} {4 \sqrt{2} \pi  B
\left\{ {\lambda_{\rm eff} [\theta(T_0)] \over 4 g^2 } \right\} \over
45 g} > 1 \, .
\ee

The last point to be discussed is the determination of the value of
$\lambda_{\rm eff} [\theta(T_0)]$ to be plugged into
eq.~(\ref{erre}). The $B$-function we use is taken from
ref.~[\ref{km}], where the sphaleron energy was computed using the
zero-temperature `Mexican-hat' potential, $V = \frac{\lambda}{4}
(\phi^2-v^2)^2$. The sphaleron energy at finite temperature was
computed in ref.~[\ref{belgians}], where it was proven that it scales
like $v(T)$, i.e.
\begin{equation}
\label{aaa}
E^{SM}_{sph}(T)=E^{SM}_{sph}(0) \; \frac{v(T)}{v} \, ,
\end{equation}
with great accuracy. Therefore, to determine the value of
$\lambda_{\rm eff} [\theta(T_0)]$ we have fitted $V_{\rm
eff}(\phi,T_0)$,
as given by eq.~(\ref{total}), to the appropriate approximate
expression,
\begin{equation}
\label{bbb}
V_{\rm eff}(\phi,T_0) \simeq \frac{1}{4} \lambda_{\rm
eff}[\theta(T_0)]
[\phi^2-v^2(T_0)]^2 + {\rm field \!\! - \!\! independent \; terms} \,
,
\end{equation}
where the field-independent terms are just to take care of the
different normalizations of the left- and right-hand sides. The value
of $\lambda_{\rm eff}$ obtained from (\ref{bbb}),
\begin{equation}
\label{ccc}
\lambda_{\rm eff}[\theta(T_0)]=4 \;
\frac{V_{\rm eff}(0,T_0)-V_{\rm eff}[v(T_0),T_0]} {v^4(T_0)} \, ,
\end{equation}
where all quantities on the right-hand side are calculated
numerically
from the potential of eq.~(\ref{total}), is then plugged into
eq.~(\ref{erre}) to obtain our bounds. We have explicitly checked the
quality of the fit in eq.~(\ref{bbb}), finding an agreement that is
more than adequate for our purposes.

Our numerical results are summarized in fig.~2, in the $\matb$ plane
and for two representative values of the top quark mass: $m_t = 130
\gev$ (fig.~2a) and $m_t = 170 \gev$ (fig.~2b). In each case, the
values of the remaining free parameters have been chosen in order
to maximize the strength of the phase transition, given the
experimental constraints on the top-stop sector. Notice that
arbitrarily small values
of $m_{U_3}$ cannot be excluded on general grounds, even if they are
disfavoured by model calculations. Also, we have explicitly checked
that,
as in ref.~[\ref{eqz3}], mixing effects in the stop mass matrix
always
worsen the case. In fig.~2, solid lines correspond to contours of
constant $R$: one can see that the requirement
of large values of $R$ favours small
$\tb$ and $m_A \gg m_Z$. The thick solid line corresponds to the
limits coming from Higgs searches at LEP: for our parameter choices,
the
allowed regions correspond to large $\tb$ and/or $m_A \gg m_Z$. For
reference, contours of constant $m_h$ (in GeV) have also been plotted
as dashed lines. One can see that, even for third-generation
squarks as
light as allowed by all phenomenological constraints,
only a very small globally allowed region can exist
in the $\matb$ plane, and that the most favourable
situation
is the one already discussed in ref.~[\ref{eqz3}]. More precisely,
the region that is still marginally allowed corresponds to $m_A \gg
m_Z$, $\tb \sim 2$, stop and sbottom sectors as light as otherwise
allowed, a heavy top, and a light Higgs boson with SM-like properties
and
mass $m_h \sim 65 \gev$, just above the present experimental limit.
A less conservative interpretation of the limits from precision
measurements, the inclusion of some theoretically motivated
constraints on the model parameters, or a few GeV improvement in the
SM Higgs mass limit, would each be enough to fully exclude
electroweak baryogenesis in the MSSM.

\vspace{1cm}
{\bf 7.}
In summary, our analysis of the full $\matb$ parameter space extends
and confirms the results of ref.~[\ref{eqz3}]: in  the region of the
MSSM parameter space allowed by the present experimental constraints,
there is very little room for fulfilling the constraint (\ref{erre}),
which is a necessary condition for electroweak baryogenesis. To put
these results in a clearer perspective,  some final comments on
possible ways out are in order.

First, one could think of relaxing the constraint $\tb \ge 1$
(and the corresponding LEP bounds), which is usually
motivated by the theoretical assumption of universal soft Higgs
masses at the SUSY-GUT scale, $M_U \sim 10^{16} \gev$. The
possibility of $\tb < 1$, however, is incompatible with a heavy top
quark, since, for $m_t \simgt 130 \gev$ and supersymmetric particle
masses of order $m_Z$, the running top Yukawa coupling would become
non-perturbative at scales smaller than $M_U$: such a possibility is
strongly disfavoured by the successful predictions of the low-energy
gauge couplings in SUSY GUTs.

A second possibility is that large non-perturbative effects,
neglected
by conventional calculational techniques, modify the predicted values
of the sphaleron energy and/or of $v(T_0) / T_0$ (for recent
suggestions
along this line, see [\ref{nonpert}]). We do not see strong physical
arguments to favour this, but we admit that it cannot be rigorously
excluded. Perhaps alternative approaches to the electroweak phase
transition [\ref{altern}]
could help clarify this point in the future.

Barring the above-mentioned possibilities, one could still try to
rescue electroweak baryogenesis by further enlarging the MSSM Higgs
sector, for example by introducing an extra singlet. Supersymmetric
models with singlets
and non-supersymmetric models, however, develop dangerous
instabilities if coupled to the superheavy sector of an underlying
unified theory. It might well be that baryogenesis has to be
described
by physics at a scale larger than the electroweak one.

\section*{Acknowledgements}
We would like to thank J.-F.~Grivaz for helping us to understand the
LEP
limits on the MSSM Higgs bosons. One of us (F.Z.) would like to thank
IEM-CSIC for the warm hospitality during the final phase of this
work. Another of us (J.R.E.) wants to thank the SCIPP for the
warm hospitality during part of this work.
\newpage
\section*{References}
\begin{enumerate}
\item
\label{reviews}
For reviews and references see, e.g.:
\\
M.E.~Shaposhnikov, in `Proceedings of the 1991 Summer School in High
Energy Physics and Cosmology', Trieste, 17 June--9 August 1991,
E.~Gava et al., eds. (World Scientific, Singapore, 1992), Vol.~1,
p.~338;
\\
A.D.~Dolgov, Phys. Rep. 222 (1992) 309;
\\
A.G.~Cohen, D.B.~Kaplan and A.E.~Nelson, Ann. Rev. Nucl. Part. Sci.
43 (1993) 27.
\item
\label{farsha}
G.~Farrar and M.E.~Shaposhnikov, Phys. Rev. Lett. 70 (1993) 2833 +
(E) 71 (1993) 210 and preprint CERN-TH.6734/93.
\item
\label{gavela}
M.B.~Gavela, P.~Hern\'andez, J.~Orloff and O.~P\`ene, preprint
CERN-TH.7081/93.
\item
\label{coignet}
See, e.g.:
G.~Coignet, Plenary talk at the XVI International Symposium on
Lepton-Photon Interactions, Cornell University, Ithaca, New York,
10--15 August 1993, to appear in the Proceedings, and references
therein.
\item
\label{limits}
M.E.~Carrington, Phys. Rev. D45 (1992) 2933;
\\
M.~Dine, R.G.~Leigh, P.~Huet, A.~Linde and D.~Linde, Phys. Lett. B283
(1992) 319 and Phys. Rev. D46 (1992) 550;
\\
P.~Arnold, Phys. Rev. D46 (1992) 2628;
\\
J.R.~Espinosa, M.~Quir\'os and F.~Zwirner, Phys. Lett. B314 (1993)
206;
\\
P.~Arnold and O.~Espinosa, Phys. Rev. D47 (1993) 3546;
\\
W.~Buchm\"{u}ller, Z.~Fodor, T.~Helbig and D.~Walliser, preprint DESY
93-021.
\item
\label{cn}
A.G.~Cohen and A.E.~Nelson, Phys. Lett. B297 (1992) 111.
\item
\label{eqz3}
J.R.~Espinosa, M.~Quir\'os and F.~Zwirner, Phys. Lett. B307 (1993)
106.
\item
\label{previous}
G.F.~Giudice, Phys. Rev. D45 (1992) 3177;
\\
S.~Myint, Phys. Lett. B287 (1992) 325.
\item
\label{cpspo}
D.~Comelli and M.~Pietroni, Phys. Lett. B306 (1993) 67;
\\
J.R.~Espinosa, J.M.~Moreno and  M.~Quir\'os, Madrid preprint
IEM-FT-76/93,
to appear in Physics Letters B (1993).
\item
\label{erz2}
J.~Ellis, G.~Ridolfi and F.~Zwirner, Phys. Lett. B262 (1991) 477.
\item
\label{altarelli}
See, e.g.: G.~Altarelli, Plenary talk given at the International
Europhysics Conference on High Energy Physics, Marseille, 22--28 July
1993, preprint CERN-TH.7045/93, to appear in the Proceedings, and
references therein.
\item
\label{abc}
R.~Barbieri, M.~Frigeni and F.~Caravaglios, Phys. Lett. B279 (1992)
169;
\\
G.~Altarelli, R.~Barbieri and F.~Caravaglios, Nucl. Phys. B405 (1993)
3, preprint CERN-TH.6859/93 and Phys. Lett. B314 (1993) 357;
\\
J.~Ellis, G.L.~Fogli and E.~Lisi, Phys. Lett. B285 (1992) 238,
B286 (1992) 85 and Nucl. Phys. B393 (1993) 3.
\item
\label{rhosusy}
L.~Alvarez-Gaum\'e, J.~Polchinski and M.~Wise, Nucl. Phys. B221
(1983) 495;
\\
R.~Barbieri and L.~Maiani, Nucl. Phys. B224 (1983) 32;
\\
C.S.~Lim, T.~Inami and N.~Sakai, Phys. Rev. D29 (1984) 1488.
\item
\label{pioneer}
Y.~Okada, M.~Yamaguchi and T.~Yanagida, Prog. Theor. Phys. Lett.
85 (1991) 1 and Phys. Lett. B262 (1991) 54;
\\
J.~Ellis, G.~Ridolfi and F.~Zwirner, Phys. Lett. B257 (1991) 83;
\\
H.E.~Haber and R.~Hempfling, Phys. Rev. Lett. 66 (1991) 1815;
\\
R.~Barbieri and M.~Frigeni, Phys. Lett. B258 (1991) 395.
\item
\label{shapo}
M.~Shaposhnikov, JETP Lett. 44 (1986) 465; Nucl. Phys. B287 (1987)
757 and B299 (1988) 797.
\item
\label{two}
A.~Bochkarev, S.~Kuzmin and M.~Shaposhnikov, Phys. Rev. D43 (1991)
369
and Phys. Lett. B244 (1990) 275;
\\
B.~Kastening, R.D.~Peccei and X.~Zhang, Phys. Lett. B266 (1991) 413.
\item
\label{km}
F.R.~Klinkhamer and N.S.~Manton, Phys. Rev. D30 (1984) 2212.
\item
\label{mixing}
J.~Kunz, B.~Kleihaus and Y.~Brihaye, Phys. Rev. D46 (1992) 3587.
\item
\label{belgians}
S.~Braibant, Y.~Brihaye and J.~Kunz, Utrecht preprint THU-93/01.
\item
\label{nonpert}
K.~Kajantie, K.~Rummukainen and M.~Shaposhnikov, Nucl. Phys. B407
(1993) 356;
\\
M.~Shaposhnikov, Phys. Lett. B316 (1993) 112.
\item
\label{altern}
B.~Bunk, E.M.~Ilgenfritz, J.~Kripfganz and A.~Schiller, Phys. Lett.
B284 (1992) 371 and Nucl. Phys. B403 (1993) 453; \\ J. March-Russell,
Phys. Lett. B296 (1992) 364; \\ H.~Meyer-Ortmanns and A.~Patk\'os,
Phys. Lett. B297 (1993) 331; \\ A.~Jakov\'ac and A.~Patk\'os,
Z.~Phys. C60 (1993) 361; \\ N.~Tetradis and C.~Wetterich, Nucl. Phys.
B398 (1993) 659 and preprint DESY-93-128; \\ P.~Arnold
and L.G.~Yaffe, University of Washington preprint UW/PT-93-24.
\end{enumerate}
\newpage
\section*{Figure captions}
\begin{itemize}
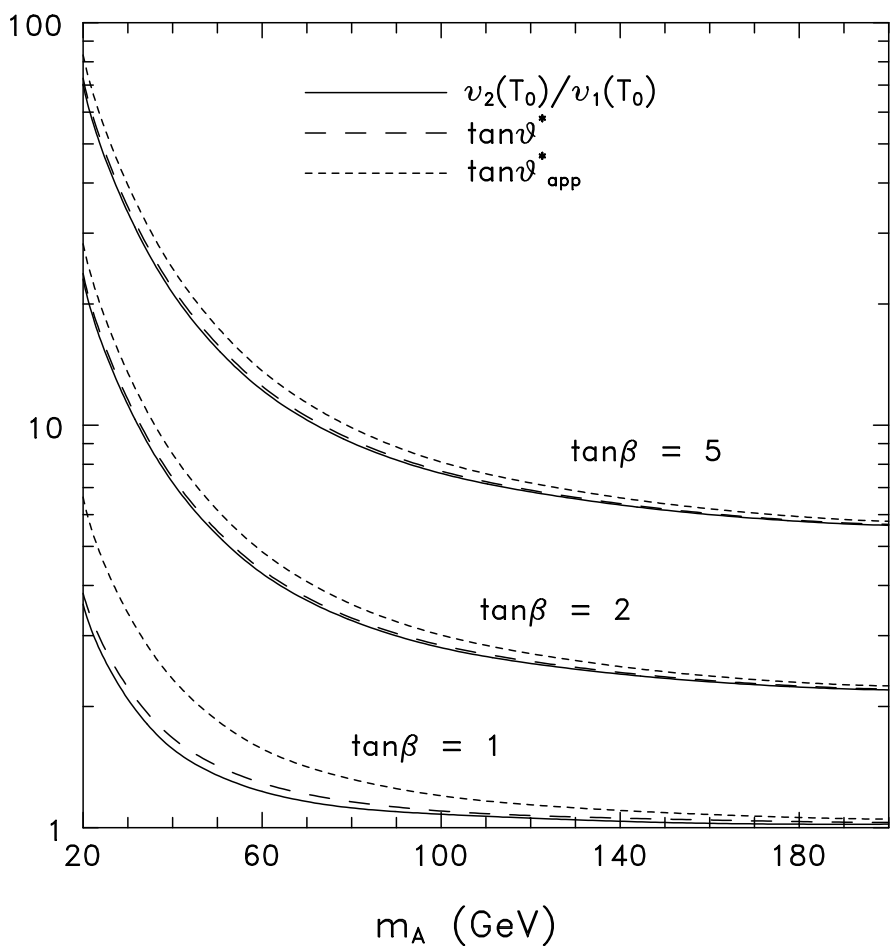
\item[Fig.1:]
The quantities $\tan \theta (T_0)$ (solid lines), $\tan \theta^*$
(dashed lines) and $\tan \tsapp$ (dotted lines), as functions of
$m_A$, for $\tan \beta = 1,2,5$ and the representative parameter
choice $m_t=150 \gev$, $m_{Q_3}=m_{U_3}=1 \tev$, $A_t = \mu = 0$.
\item[Fig.2:]
Contours of $R$ in the $\matb$ plane, for the parameter choices:
a) $m_t = 130 \gev$, $m_{Q_3} = 50 \gev$, $m_{U_3} = 0$
($m_{\tilde{t}} \sim~130 \gev,\ m_{\tilde{b}_L} \sim~50 \gev$),
$ A_t = \mu = 0$;
b)~$m_t = 170 \gev$, $m_{Q_3} = 280 \gev$, $m_{U_3} = 0$
($m_{\tilde{t}_L} \sim~330 \gev$, $m_{\tilde{t}_R} \sim~170 \gev$,
$m_{\tilde{b}_L} \sim~280 \gev$),
$A_t = \mu = 0$.
The region excluded by Higgs searches at LEP is delimited by the
thick solid line. For reference, contours of constant $m_h$ (in GeV)
are also represented as dashed lines.
\end{itemize}
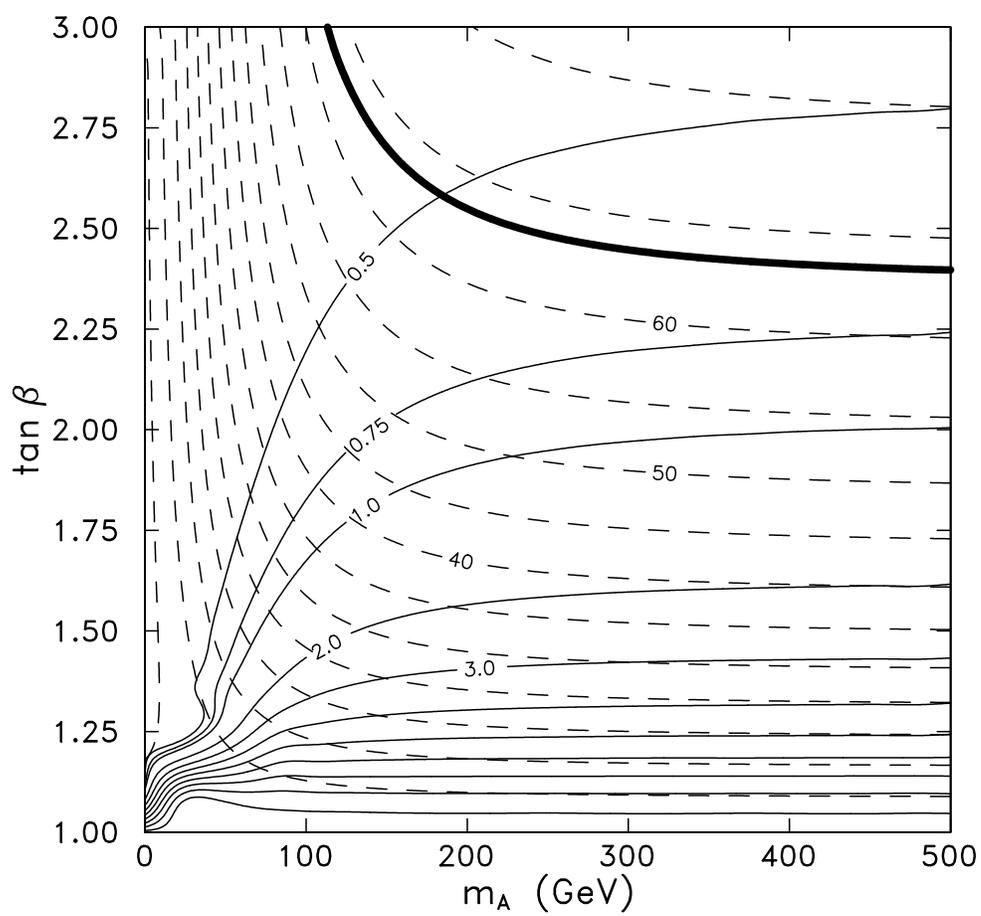
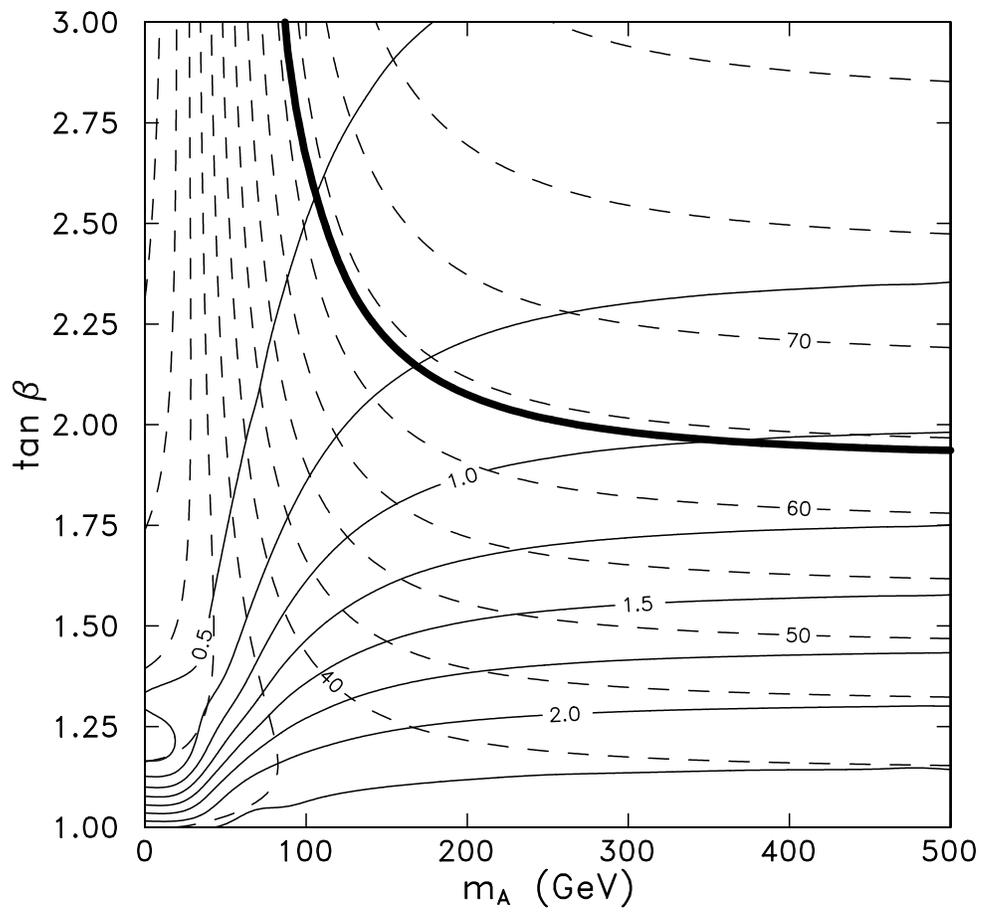
\end{document}